\documentclass[10pt, conference]{IEEEtran}
\IEEEoverridecommandlockouts
\usepackage{cite}
\usepackage{amsmath,amssymb,amsfonts}
\usepackage{algorithmic}
\usepackage{graphicx}
\usepackage{textcomp}
\usepackage{xcolor}
\usepackage{units}
\usepackage{multirow}
\usepackage{colortbl}
\usepackage{soul}
\usepackage[left=0.680in, right=0.6in,top=0.7in,bottom=1in]{geometry}

\usepackage{amsthm}
\usepackage{thmtools, thm-restate}
\usepackage{algorithm}
\newcommand{\A}{\mathbf{A}}
\renewcommand{\a}{\mathbf{a}}

\renewcommand{\c}{\mathbf{c}}

\newcommand{\C}{\mathbf{C}}
\newcommand{\Cb}{\mathbb{C}}

\newcommand{\G}{\mathbf{G}}
\newcommand{\g}{\mathbf{g}}
\newcommand{\Hb}{\mathbf{H}}
\newcommand{\h}{\mathbf{h}}

\newcommand{\I}{\mathbf{I}}

\newcommand{\n}{\mathbf{n}}
\newcommand{\p}{\mathbf{p}}

\newcommand{\Rb}{\mathbb{R}}
\renewcommand{\r}{\mathbf{r}}

\newcommand{\s}{\mathbf{s}}

\renewcommand{\v}{\mathbf{v}}

\newcommand{\X}{\mathbf{X}}

\newcommand{\mc}[1]{\mathcal{#1}}

\newcommand{\The}{\mathbf{\Theta}}
\newcommand{\theb}{\boldsymbol{\theta}}
\newcommand{\phib}{\boldsymbol{\phi}}

\newcommand{\alg}{$\mathsf{PAPA}~$}

\def\BibTeX{{\rm B\kern-.05em{\sc i\kern-.025em b}\kern-.08em
    T\kern-.1667em\lower.7ex\hbox{E}\kern-.125emX}}
\IEEEoverridecommandlockouts\IEEEpubid{\makebox[\columnwidth]{ 978-1-6654-3540-6/22~\copyright~2022 IEEE \hfill} \hspace{\columnsep}\makebox[\columnwidth]{ }} 
\begin{document}

%\title{Iterative Power Control Algorithm for Distributed Reconfigurable Intelligent Surfaces Assisted Wireless Networks
\title{Iterative Power Control for Wireless Networks with Distributed Reconfigurable Intelligent Surfaces 
\thanks{This work is supported in part by NSF AI-EDGE Institute (2112471).}
}

% \and
% \and
% % \and
% % \IEEEauthorblockN{4\textsuperscript{th} Given Name Surname}
% % \IEEEauthorblockA{\textit{dept. name of organization (of Aff.)} \\
% % \textit{name of organization (of Aff.)}\\
% % City, Country \\
% % email address or ORCID}
% % \and
% % \IEEEauthorblockN{5\textsuperscript{th} Given Name Surname}
% % \IEEEauthorblockA{\textit{dept. name of organization (of Aff.)} \\
% % \textit{name of organization (of Aff.)}\\
% % City, Country \\
% % email address or ORCID}
% % \and
% % \IEEEauthorblockN{6\textsuperscript{th} Given Name Surname}
% % \IEEEauthorblockA{\textit{dept. name of organization (of Aff.)} \\
% % \textit{name of organization (of Aff.)}\\
% % City, Country \\
% % email address or ORCID}
% }


\author{Jiayu Mao and Aylin Yener
\\ INSPIRE@OhioState Research Center %Information and Networked Systems Powered by Innovation and Research in Engineering
\\Dept. of Electrical and Computer Engineering
\\ The Ohio State University
\\ mao.518@osu.edu, yener@ece.osu.edu 
}

\newgeometry{left=0.680in, right=0.60in,top=0.812in,bottom=0.95in}
\maketitle

% !TEX root = main.tex

\begin{abstract}

Reconfigurable Intelligent Surfaces (RIS) are a new paradigm which, with judicious deployment and alignment, can enable more favorable propagation environments and better wireless network design. As such, they can offer a number of potential benefits for next generation wireless systems including improved coverage, better interference management and even security. In this paper, we consider an uplink next generation wireless system where each user is assisted with an RIS. We study the uplink power control problem in this distributed RIS-assisted wireless network.
Specifically, we aim to minimize total uplink transmit power of all the users subject to each user's reliable communication requirements at the base station by a joint design of power, receiver filter and RIS phase matrices. We propose an iterative power control algorithm, combined with a successive convex approximation technique to solve the problem with non-convex phase constraints. Numerical results illustrate that distributed RIS assistance leads to uplink power savings when direct links are weak.
\end{abstract}

%Reconfigurable Intelligent Surface (RIS) has recently been considered a new promising technique for 5G and beyond. It can extend the communication coverage, especially when the direct LoS path is blocked. In this paper, we consider power minimization in a distributed RISs-assisted wireless network, where each user is equipped with one RIS nearby. We aim to minimize total uplink transmit power under SINR requirements by a joint design of power, receiver filter and RIS phase matrices. We propose an iterative power control algorithm, combined with successive convex approximation technique to solve the problem with non-convex phase constraints. The simulation results illustrate that distributed RISs save the uplink power when direct links are weak.


\section{Introduction}
\label{sec:intro}
%To date, while the deployment of the fifth generation of wireless communication (5G) systems is ongoing, the inherent limitations of such systems are continuously exposed.  Meanwhile, 
Future wireless networks continue to demand more than the current generation (5G) can deliver. In particular, emerging applications include those that require ultra high data rates, high reliability and spectral efficiency simultaneously \cite{saad2019vision}. To that end, new physical layer innovations are needed for 6G. Reconfigurable intelligent surfaces (RIS) -- also known as intelligent reflecting surfaces (IRS)~\cite{wu2019intelligent,wu2019towards,wu2021intelligent} -- can facilitate such innovations. %wu2019towards,
An RIS is a planar meta-surface that contains a large number of passive reflecting elements, each of which is capable to independently adjust the reflecting phase and (possibly) amplitude of the incident signal~\cite{wu2019intelligent}. By judicious deployment of RISs, propagation environments of wireless signals can be steered towards desired channel realizations and/or distributions. A main advantage of RIS is its low-cost implementation. 
Specifically, compared to massive MIMO, RIS deployment results in remarkably smaller number of the transmitter antennas when achieving a target beamforming~\cite{hu2018beyond}.
In addition, an RIS operates (effectively) in full-duplex (FD) mode without requiring active radio-frequency (RF) chains and generating any antenna noise amplification or self-interference, and is naturally advantageous over relay assisted communications \cite{chen2008distributed}.

%Due to the above fascinating merits, RIS is naturally favorable to be integrated into wireless networks as an auxiliary device to enhance spectral and energy efficiency. In particular, 

The topic of this paper is power efficient transmissions via RIS assistance. To date, energy and resource efficiencies of RIS-assisted wireless networks have been studied in various works~\cite{zhao2020energy,wu2020jointpower,guo2021joint, fu2021reconfigurable}.
Broadly, these efforts can be categorized into two.
The first line of research focuses on maximizing energy efficiency of the communication system. 
For example, in \cite{zhao2020energy}, the system energy efficiency is maximized by jointly optimizing the RIS phase matrix and the transmit power.
The second line of work focuses on minimizing the total transmit power.
In \cite{wu2020jointpower}, the transmit power of the base station for a downlink MISO system is minimized through the optimization of RIS phase shifts and the transmit precoders.
Reference \cite{guo2021joint} considers a RIS-aided downlink mmWave communication system and a penalty-based algorithm is proposed to minimize total transmit power under SINR constraints. 
In \cite{fu2021reconfigurable}, RISs are used to empower downlink NOMA systems. An alternating optimization framework is proposed to minimize total power consumption with rate requirements at the receivers.

RISs in power minimization in the uplink is more recent %While most existing works investigate downlink RIS-aided systems, research on total power minimization of the uplink RIS-assisted communications remains in its infancy
~\cite{zhou2021joint,wang2020power,wang2022simultaneous,cao2021delay,wu2022energy,ma2021power,liu2020intelligent}. 
%In particular, on the one hand, several works have studied different uplink RIS-assisted systems~\cite{zhou2021joint,wang2020power,wang2022simultaneous}.
In \cite{zhou2021joint}, an RIS-aided IoT sensor network with sleep and active nodes is considered.
\cite{wang2020power} discusses an RIS-aided two-cell NOMA network.
A STAR-RIS assisted full-duplex communication system is studied in \cite{wang2022simultaneous}.
%To date, different optimization techniques are applied to solve the uplink power minimization problem in RIS-assisted systems.
%Specifically, there are two main difficulties in solving the above problem~\cite{pan2021overview}. 
%First, the overall problem is non-convex in nature due to the unit-modulus constraint of RIS phase shifts.
%Second, the beamforming vectors and RIS phase design are usually coupled in the constraints (e.g., SINR or rate requirements).
In \cite{cao2021delay}, an alternating optimization method is developed to decompose the joint optimization problem of powers and the RIS phases. The RIS phase configuration is solved by the alternating direction method of multipliers. In \cite{wu2022energy}, a Riemannian conjugate gradient-based alternating optimization scheme is proposed, which exploits the Riemannian manifold structure to deal with the unit-modulus phase constraint. In \cite{ma2021power}, semi-definite relaxation is applied for the phase shift optimization problem.
In \cite{liu2020intelligent}, nonlinear equality constrained alternative direction method of multipliers based solutions are utilized.

Different than existing works that consider uplink power minimization assisted by a single RIS, in this paper, we consider uplink power minimization in a model where {\it multiple} RIS units distributed in the network assist end users to communicate to the base station. In particular, we envision systems in which as many RISs as users are deployed and consider two variations of the system model, one in which each user has a personal RIS whose reception from its user is the dominant one (for example each RIS deployed is in a residence in which the transmitter is located), and another, a fully connected network, where each RIS receives and reflects signals transmitted by all users (for example when RISs are deployed in a public space with the intention of assisting individual users). We formulate the joint transmit power and RIS phase optimization problem subject to SINR constraints and propose an iterative power control (and phase control) algorithm for its solution. Specifically, we jointly design the receiver filters at the multiple antenna base station, transmit power levels at each user and the RIS phase shifts in a distributed manner. To tackle the non-convex optimization challenges, we first apply alternating optimization to decouple the variables, then we exploit the successive convex approximation (SCA)~\cite{pan2021overview} for phase updates. A point we wish to stress is that existing works for power minimization in RIS assisted systems rely on  centralized optimization algorithms. By contrast, by design, the proposed algorithm in this paper is distributed in nature similar to the power control, receiver and transmitter optimization approaches for non-orthogonal uplinks in systems that pre-date RISs \cite{yener2001inter}.
%The SCA method is another widely used technique to deal with the unit-modulus phase constraint~\cite{pan2021overview}, which iteratively approximates a difficult problem into several simpler subproblems to solve. 
Numerical results demonstrate the convergence of the proposed iterative algorithm and the power savings. %brought on the  the proposed approach saves uplink power when direct links are weak.
%In other words, the RIS is controlled by the base station.  However, in practice, RIS may locate in a large distance from the base station, which results in a difficulty to achieve a low-latency phase design.  Moreover, the computation resources of users in wireless networks are not fully explored in these existing works.These limitations motivate us to pursue a distributed system design that utilizes the edges intelligence to control RISs and save the uplink power under QoS constraints.

%We emphasise that our joint power control and phase design algorithm~\alg operates in a fully distributed manner, which differences us from the literature. 

%Our main contributions are summarized as follows:
%\vspace{-.05in}
%\begin{list}{\labelitemi}{\leftmargin=1em \itemindent=-0.5em \itemsep=.2em}
%\item We propose a power control algorithm~\alg that jointly designs receiver structure, power and phase shifts for distributed RISs-assisted wireless communication systems.
%\item Our proposed algorithm~\alg operates in a fully distributed fashion. In each iteration, each user can compute all the updates independently, which explores the computation resources of edge devices.
%\item We demonstrate the effectiveness of~\alg algorithm in numerical results. We show that the deployment of distributed RISs can save uplink power when the direct links are weak.
%\end{list}

\section{System Model}
\label{sec: prelim}
\begin{figure}[t]
    \centering
    \includegraphics[scale=0.26]{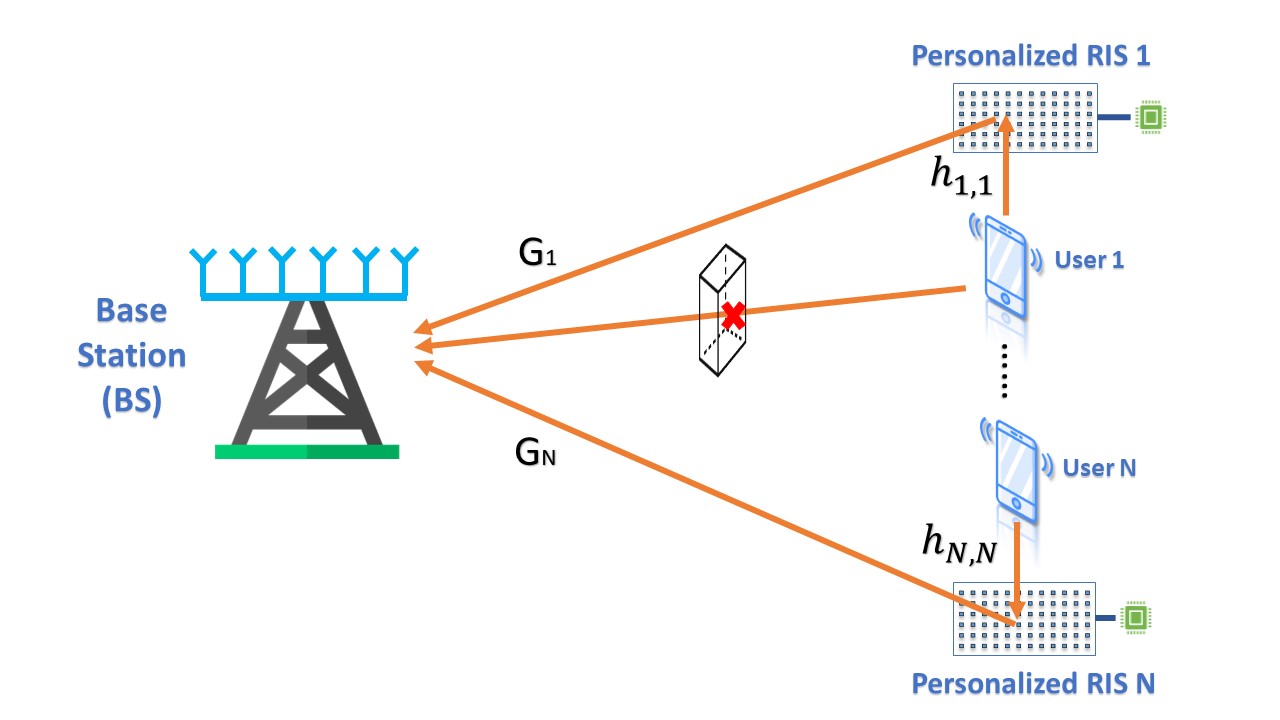}
    \caption{Distributed personal RIS-assisted multiuser communications.}
    \label{fig:sys_noi}
\end{figure}
\begin{figure}[t]
    \centering
    \includegraphics[scale=0.26]{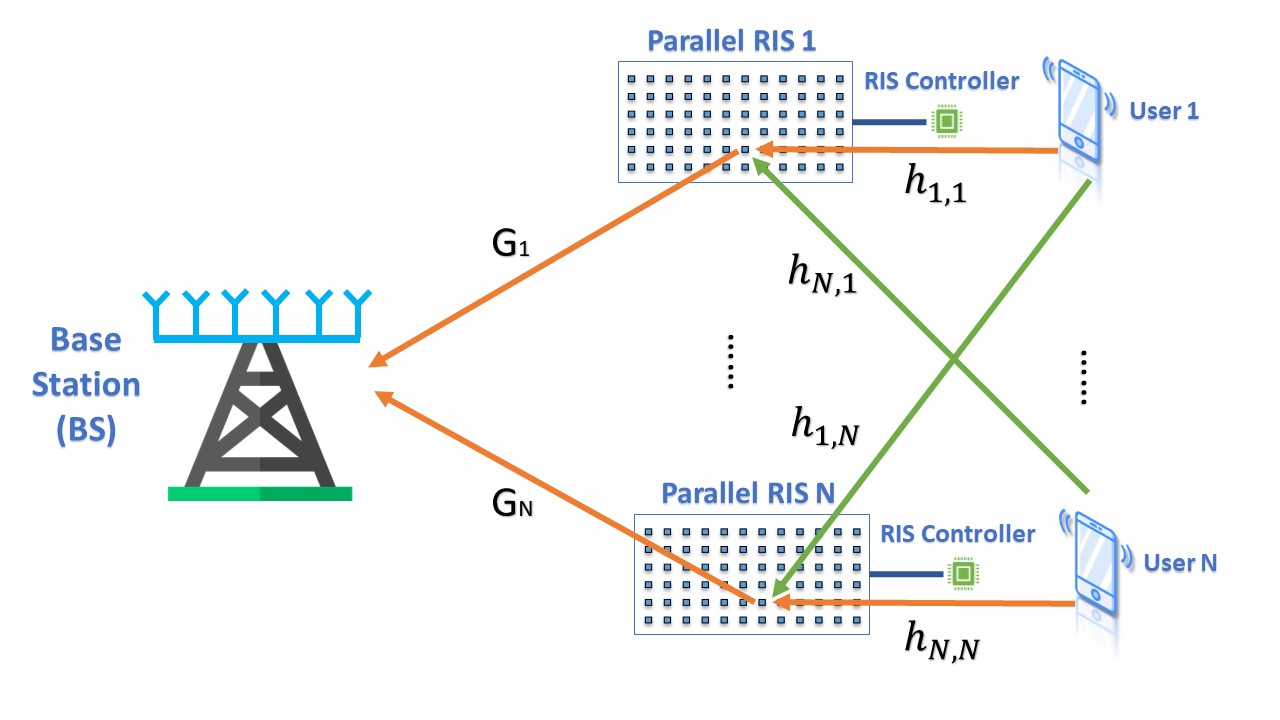}
    \caption{Distributed parallel RIS-assisted multiuser communications.}
    \label{fig:sys_inter}
\end{figure}
We consider two models for RIS-assisted uplinks, the personal RIS-assisted model and the parallel RIS-assisted model, as shown in Fig.~\ref{fig:sys_noi} and Fig.~\ref{fig:sys_inter}, respectively. 
Each system consists of $N$ single antenna users, $N$ RISs and one base station (BS) equipped with $M$ antennas.
Each user is assisted with an RIS which contains $K$ passive reflecting elements. 
We consider the scenario where is no direct link available between BS and users, which renders RIS-assistance essential.  %as the BS and users are located far apart from each other and direct links are blocked, or every direct link experiences large path attenuation that is negligible compared to the RIS assisted link. 

For user $i$, $p_i$ %to represents  
represents its transmitted power, and $b_i$ represents its transmitted bit.
The equivalent baseband channels from user $i$ to RIS $j$, from RIS $j$ to BS are represented by $\h_{i,j} \in \Cb^K$, $\G_j = [\g_{j,1},...,\g_{j,M}]^T \in \Cb^{M \times K}$, respectively, where $\g_{j,m}$ denotes the channel between RIS $j$ and $m$-th antenna of BS. The channel model is %We use a free space path loss channel model
\begin{equation}
    \h_{i,j}= \frac{1}{d_{i,j}^{\alpha}} \a(\theta_l, \phi_l),
\end{equation}
where $d_{i,j}$ is distance between user $i$ and RIS $j$, $\alpha$ is pathloss exponent, $\a(\theta_l, \phi_l) \in \Cb^K$ represents phase shifts at the azimuth $\theta_l$ and elevation $\phi_l$ angles of arrival~\cite{erpek2021autoencoder}.
Similarly, $\g_{j,m}$, $\forall m \in [M]$ follows same definition.
We assume that channel coefficients remain constant during one communication period,
both BS and users have perfect channel state information. 
We define $\The_j$ as phase matrix of RIS $j$, where $\The_j = diag (\theta_{j,1}, \theta_{j,2},..., \theta_{j,K})$ is a diagonal matrix, $\theta_{j,k} = e^{j \phi_{j,k}}$ is the phase of the $k$-th reflecting element. 
The signal transmitted to RIS will be passively reflected to BS, with phase multiplication to the original incident signal. 

We first consider the personal RIS-assisted model, i.e., each user can only transmit through its own RIS in Fig.~\ref{fig:sys_noi}.
This model is suitable when the channels between users and other RISs are weak. 
The received signal at the BS is expressed as\footnote{Without loss of generality, we consider synchronous models.} 
\begin{equation}
    \r = \sum_{i=1}^{N} \G_i \The_i \h_{i,i} \sqrt{p}_i b_i + \n,
\end{equation}
where $\n$ is additive white Gaussian noise with variance $\sigma^2$.
Let $\c_j \in \Cb^M$ represent the receiver filter for user $j$ at base station. The receiver filter output of user $j$ is
\begin{equation}
    y_j = \c_j^H \sum_{i=1}^{N}\G_i \The_i \h_{i,i} \sqrt{p}_i b_i + \Tilde{n}_j,
\end{equation}
where $\Tilde{n}_j = \c_j^H \n$ is Gaussian random variable with zero mean and variance $\sigma^2 \c_j^H \c_j$. 
Accordingly, the signal-to-interference-plus-noise ratio (SINR) of user $j$ is given by
\begin{equation}
    \gamma_j = \frac{|\c_j^H\G_j \The_j \h_{j,j} \sqrt{p}_j|^2}{\sum_{i \neq j}^{N} |\c_j^H \G_i \The_i \h_{i,i} \sqrt{p}_i|^2 + \sigma^2 \c_j^H \c_j}.
\end{equation}

Next, we take interference between users and other RISs into account and call it parallel RIS model, as illustrated in Fig.~\ref{fig:sys_inter}. Similarly, the received signal can be written as 
\begin{equation}
    \tilde{\r} = \sum_{i=1}^{N} \G_i \The_i \sum_{j=1}^{N} \h_{i,j} \sqrt{p}_i b_i + \n,
\end{equation}
where each user transmits through all RISs. User $j$'s SINR is 
\begin{equation}
    {\tilde{\gamma}}_j = \frac{|\c_j^H \sum_{i=1}^{N}\G_i \The_i \h_{j,i} \sqrt{p}_j|^2}{|\c_j^H \sum_{n \neq j}^{N} \sum_{i=1}^{N} \G_i \The_i \h_{n,i} \sqrt{p}_n|^2 + \sigma^2 \c_j^H \c_j}.    
\end{equation}
%YENER EDITED TILL HERE 7pm May 21
\section{Power Minimization for Uplink Transmission} 
\label{sec: prob}
We aim to minimize the total transmitted power of all users by jointly designing the receiver filters $c_i$, powers $p_i$ and phase matrices $\Theta_i$ of RISs for $i = 1,...,N$ users, while ensuring each user $i$ reaches its SINR target (quality of service requirement), denoted by $\gamma_i^*$. For both models described in Section~\ref{sec: prelim}, we shall denote the SINR as $\gamma_i \geq \gamma_i^*$ for $i = 1,...,N$, the nuances in the two respective algorithms will be explained in the sequel. Thus, the optimization problem is formulated as 
\begin{equation}
    \begin{aligned} \label{prob:P1}
         \mc{P}_1: \mathop{min}\limits_{\c_i, \p, \The_i}& &&\sum_{i=1}^{N} p_i \\
        s.t.& &&  \gamma_i \geq \gamma_i^*, \quad i=1,...,N,\\
        & && p_i \geq 0, \qquad i=1,...,N,\\
        & && \c_i \in \Cb^M, \quad  i=1,...,N, \\
        & && |\theta_{i,k}|=1,  \quad i=1,...,N, k=1,...,K.
    \end{aligned}
\end{equation}
Note that in~(\ref{prob:P1}), the unit-modulus phase constraint of RIS $i$ is represented by $|\theta_{i,k}|=1, \forall k \in [K]$.
Due to the non-convexity of this unit-modulus constraint, $\mc{P}_1$ is a non-convex optimization problem. 
In addition, receiver filter, power and phase matrices are coupled. % together in SINR requirements. Therefore, the non-convex property and the coupled structure make $\mc{P}_1$ difficult to solve.

To handle these challenges, we employ alternating minimization \cite{yener2001inter} to decouple the optimization variables combined with the successive convex approximation technique to deal with the non-convex constraint. Alternating optimization is essentially a two-block version of block coordinate descent method, the basic idea is to fix the value of all but one of the variables and update that variable and iterate over different variables until convergence~\cite{xu2013block}. We note that this approach is by now a common method in RIS-assisted system to address the coupled structure~\cite{pan2021overview}. In particular, the joint optimization problem is decomposed into two: one is the beamforming design problem, and the other is the phase design problem.
As such, we are able to propose an iterative algorithm to solve $\mc{P}_1$. We first solve $\mc{P}_1$ for the personalized RIS model, then extend the solution to the parallel RIS case in section~\ref{sec: alg}. We term the overall iterative algorithm \underline{P}hase \underline{A}ware \underline{P}ower Control \underline{A}lgorithm: \alg.

\subsection{Power Control and Filter Design}
\label{subsec:pc}
When the phase shifts $\The_i, \forall i \in [N]$ are given, the optimization problem reduces to a joint beamforming and power control problem for multiuser system, which has been extensively studied in the literature in the context of CDMA systems, for single antenna \cite{ulukus1998adaptive} and multiple antenna uplink \cite{yener2001inter}.

We can thus reformulate the problem to its following equivalent:
\begin{equation}
    \begin{aligned} \label{prob:P2}
        & \mathop{min}\limits_{\p} \quad \sum_{i=1}^{N} p_i \\
        &\begin{array}{ll}
        s.t. &  p_i \geq \gamma_i^*\mathop{min}\limits_{\c_i \in \Cb^M} \frac{\sum_{j \neq i}^{N} |\c_i^H \G_j \The_j \h_{j,j} |^2 p_j + \sigma^2 \c_i^H \c_i}{|\c_i^H\G_i \The_i \h_{i,i} |^2},\\ 
        & i=1,...,N,\\
        & p_i \geq 0, \quad i=1,...,N.
        \end{array}
    \end{aligned}
\end{equation}
where the outer optimization problem is defined over the power vector only, the inner optimization depends on the receiver filter \cite{ulukus1998adaptive}. 
For ease of representation, we define the effective end-to-end spatial signature as 
\begin{equation}
    \s_i = \G_i \The_i \h_{i,i}. \label{equ:spasig}
\end{equation}
Then the iterative updating rule of user $i$ in iteration $t+1$ can be expressed as \cite{ulukus1998adaptive}:
\begin{equation}
    \hat{\c}_i = \frac{\sqrt{p_i(t)}}{1 + p_i(t) \s_i^H \A_i^{-1}(\p(t)) \s_i} \A_i^{-1}(\p(t)) \s_i, \label{filter}
\end{equation}
\begin{equation}
    p_i(t+1) = \gamma_i^* \frac{\sum_{j \neq i}^{N} |\hat{\c}_i^H \s_j |^2 p_j(t) + \sigma^2 \hat{\c}_i^H \hat{\c}_i}{|\hat{\c}_i^H \s_i |^2}, \label{power}
\end{equation}
where $\A_i(\p(t)) = \sum_{j \neq i} p_j(t) \s_j \s_j^H + \sigma^2 \I$. 
Note that the filter coefficients are the MMSE filters, and the transmit power equals the corresponding interference. It has been shown in~\cite{ulukus1998adaptive} that the iterative algorithm (\ref{filter})-(\ref{power}) converges to the minimum power. Thus, when the phase shifts are fixed, filters and power vector are optimized by this iterative algorithm. 

\subsection{Phase Design}
\label{subsec:phase}
We now fix power $\p$ and filters $\c_i$ and focus on the phase design $\The_i, \forall i \in [N] $, which only relates to SINR constraints. 
Note that phase design is equivalent to design the effective end-to-end spatial signature in~(\ref{equ:spasig}).
A natural idea here is to relax the problem to optimize the effective end-to-end spatial signatures first, and then update the phase shifts to as close as possible to the optimal sequence design.

Similar to~(\ref{prob:P2}), SINR requirements can be rewritten as $N$ parallel inner optimization problems. 
To make these problems more tractable, we rewrite $\s_i = \G_i \Hb_{i,i} \theb_i$, where $\Hb_{i,i} = diag(\h_{i,i})$ is diagonal matrix, and $\theb_i = (\theta_{i,1},...,\theta_{i,K})^T$. 
Due to the unit-modulus constraint, each phase vector $\theb_i$ has norm $K$. 
Then, the above phase constraint can be equivalently written as $\|\s_i\| \leq \|\G_i\Hb_{i,i}\|\|\theb_i\| = \sqrt{K}\|\G_i\Hb_{i,i}\|$.
Therefore, we represent the inner optimization problem of user $i$ as
\begin{equation}
    \begin{aligned} \label{prob:P3}
        \mc{P}_2: & \mathop{min}\limits_{\s_i} \quad \frac{\sum_{j \neq i}^{N} |\c_i^H \s_j |^2 p_j + \sigma^2 \c_i^H \c_i}{|\c_i^H \s_i |^2} \\
        &\begin{array}{ll}
        s.t. & \|\s_i\| \leq \sqrt{K}\|\G_i\Hb_{i,i}\|, \quad i=1,...,N.
        \end{array}
    \end{aligned}
\end{equation}

The optimal effective end-to-end spatial signature update of user $i$ is found as \cite{ulukus2004iterative}:
\begin{equation}
    \s_i = \sqrt{p_i} (p_i \C \C^H + \mu_i \I_i)^{-1} \c_i, \label{sigse}
\end{equation}
where $\C = [\c_1,...,\c_N]$ is a $M \times N$ matrix containing all receiver filters, $\mu_i$ is the Lagrange multiplier chosen to satisfy the norm constraint of $\s_i$. 

Next, we consider the phase update.
To approximate the optimal end-to-end spatial signature update, we formulate a regression problem to minimize the difference between the optimal update and the effective spatial signature.
Specifically, the phase update problem of user $i$ is
\begin{equation}
    \begin{aligned} \label{prob:P4}
        \mc{P}_3: & \mathop{min}\limits_{\theb_i} \quad ||\s_i - G_i \Hb_{i,i} \theb_i||_2^2 \\
        &\begin{array}{cc}
        s.t. & |\theta_{i,k}|=1, \quad k=1,...,K.
        \end{array}
    \end{aligned}
\end{equation}

The challenge derives from the non-convex constraint of RIS elements, which can not be solved by least square solution directly.
To overcome this difficulty, we exploit the successive convex approximation (SCA) method to find a stationary solution of $\mc{P}_3$ \cite{scutari2013decomposition,guo2020weighted}.
The basic idea of SCA is to solve a difficult problem via solving a sequence of simpler surrogate problems iteratively.
In particular, each surrogate function needs to satisfy strongly convex and differentiable constraints~\cite{scutari2013decomposition}.

We define the objective function and expand it first:
\begin{equation}
\begin{array}{ll}
     f(\theb_i) & = ||\s_i - G_i \Hb_{i,i} \theb_i||_2^2   \\
     & = \s_i^H \s_i - 2 Re\{\theb_i^H \v\} + \theb_i^H \X \theb_i,
\end{array}
\end{equation}
where $\X = \Hb_{i,i}^H \G_i^H \G_i \Hb_{i,i}$, $\v = \Hb_{i,i}^H \G_i^H \s_i$. Furthermore, we replace $\theta_{i,k} = e^{j \phi_{i,k}}, \phi_{i,k} \in \Rb$. Since $\s_i$ is fixed in this problem, it is equivalent to minimize
\begin{equation}
    f_1(\phib_i) = (e^{j\phib_i})^H \X e^{j\phib_i} - 2 Re\{(e^{j\phib_i})^H \v\},
\end{equation}
where $\phib_i = (\phi_{i,1},...,\phi_{i,K})^T$.

We now implement SCA technique. We adopt the second order Taylor expansion to construct the surrogate function of $f_1(\phib_i)$ at point $\phib_i^n$ in iteration $n$:
\begin{equation}
\begin{array}{ll}
      g(\phib_i,\phib^n_i) & = f_1(\phib^n_i) + \nabla f_1(\phib^n_i)^T (\phib_i - \phib^n_i)\\
     &  + \frac{\lambda}{2} ||\phib_i - \phib^n_i||_2^2,
\end{array}
\end{equation}
where $\nabla f_1(\phib^n_i)$ is the gradient, and $\lambda$ is chosen to satisfy the constraint of the surrogate function, i.e, $g(\phib_i,\phib^n_i) \geq f_1(\phib_i) $. Then, the iterative update rule of $\phib_i$ is:
\begin{equation}
    \phib_i^{n+1} = \phib^n_i - \frac{\nabla f_1(\phib^n_i)}{\lambda}. \label{phase}
\end{equation}
After we obtain a stationary solution of phase design, we arrive at the final update of effective end-to-end spatial sequences $\s_i = \G_i \Hb_{i,i} \theb_i$.

\section{\alg and extension to Parallel-RIS setting} 
\label{sec: alg}

In this section, we summarize the proposed alternating optimization and SCA phase update for personalized RIS model in \alg, then extend it to parallel RIS scenario.
%\subsection{\alg and Convergence Discussion}
%\vspace{-1in}
\begin{algorithm}%[t!] 
    \caption{\ul{P}hase-\ul{a}ware \ul{P}ower Control \ul{A}lgorithm (PAPA).} \label{alg:pcpa} 
    \begin{algorithmic}[1]
    \STATE 
    \emph{\bf Initialization: $\p$, $\c_i$, $\theb_i, \s_i = \G_i \Hb_{i,i} \theb_i, i \in [N]$.}
    \FOR{$t=0, \dots, T-1$}
    %\STATE {Each device uploads $\x_t$ to base station.}
        \FOR{$z=0, \dots, Z-1$}
        %\STATE {Each client locally computes phase design under power constraints.} 
            \FOR{each user $i \in [N]$}
            \STATE {Update receiver filter $\c_i$ by~\eqref{filter}}; 
            \STATE {Update power $p_i$ by~\eqref{power}}; 
            \ENDFOR
        \ENDFOR
        \FOR{each user $i \in [N]$}
            \STATE {Update effective end-to-end spatial signature $\s_i$ by~\eqref{sigse}}; 
            \FOR{$j=0, \dots, J-1$}
            \STATE {Update RIS phase design $\phib_i$ by~\eqref{phase}}; 
            \ENDFOR
            \STATE {$\theb_i = e^{j\phib_i}$};
            \STATE {$\s_i = \G_i \Hb_{i,i} \theb_i$};
        \ENDFOR
    \STATE {The base station computes total transmitted power.}
    \ENDFOR
    \end{algorithmic}
\end{algorithm}

\subsection{\alg and Convergence Discussion}
As alluded to earlier, \alg is a two-stage algorithm. The algorithm is initialized to random power vector $\p$, receiver filters $\c_i$ and random phase design $\The_i , \forall i \in [N]$.
First, we fix phase shifts, update the receiver filters to be MMSE filters and the transmit powers to meet SINR requirements for the chosen filter coefficients. 
Second, we compute the optimal effective end-to-end spatial signature according to the filters and power vector, then apply SCA for parallel inner optimization problems of phase design to approximate the optimal $\s_i$. After the phase update is done, the effective end-to-end spatial signature is calculated and the alternating optimization returns to the first stage. Each stage involves two subproblems, and each subproblem requires iterative updating rule to solve. The algorithm is outlined in Algorithm~\ref{alg:pcpa}.

Unlike previous multiuser systems \cite{ulukus1998adaptive,ulukus2004iterative}, due to the physical constraints brought upon by the RISs, the effective end-to-end spatial signature $\G_i \Hb_{i,i} \theb_i$ is not guaranteed to lie in the same vector space of optimal $\s_i$, meanwhile phase design is limited by its non-convex unit-modulus constraint. By SCA, phase optimization problems can converge to a stationary solution at the second stage, and the convergence of the overall algorithm~\alg is observed in the numerical results (Section~\ref{sec: exp}). 

%A key advantage of the proposed algorithm~\alg is its fully distributed nature. %On one hand, users can update their receiver filters and power in parallel fashion, since both are independent of each other. -- I DON"T THINK THIS IS CORRECT.
%On the other hand, the optimal effective end-to-end spatial signature update only depends on power and filters and not the spatial signature of the other users, thereby users can update it independently as well. 
%Finally, the phase update of each user's RIS only happen locally, since inner phase optimization problems are independent at the second stage. 
%Thus, the proposed algorithm~\alg can be implemented in a fully distributed manner.
%Moreover, we fully exploit the computation resource of edge users and utilize the parallel fashion to speed up the convergence. 
%We emphasize that the need for distributed setting is nontrivial in the era of 6G, as we already enter a data-explosion world. 
\subsection{\alg with Parallel RISs}
Next, we discuss the extension of~\alg for the parallel RIS model.
As shown in Fig.~\ref{fig:sys_inter}, in this interference model, each user's signal is routed through all RISs to the BS.
Thus, the effective end-to-end spatial signature of user $i$ becomes
\begin{equation}
 {\tilde{\s}}_i = \sum_{j=1}^{N} \G_j \The_j \h_{i,j}   
\end{equation}
%\[{\tilde{\s}}_i = \sum_{j=1}^{N} \G_j \The_j \h_{i,j} \]. 
Again, by exchanging the position of phase matrix and channel vector, we can rewrite it as $\tilde{\s}_i = \sum_{j=1}^{N} \G_j \Hb_{i,j} \theb_{j}$. 
With this modification, \alg remains applicable for solving the power minimization problem for this parallel RIS model, while we need a slightly different inner phase optimization problem to solve.  Note that each user can only update its own RIS.
At the second stage, after the optimal end-to-end spatial signatures are obtained, users are able to start phase design in a parallel fashion. 
Define %$\s'_i = {\tilde{\s}} - \sum_{j \neq i}^{N} \G_j \Hb_{i,j} \theb_{j} (n)$, 
\begin{equation}
  \s'_i = {\tilde{\s}}_i - \sum_{j \neq i}^{N} \G_j \Hb_{i,j} \theb_{j} (n)  
\end{equation}
then the objective function of inner optimization becomes $||\s'_i - G_i \Hb_{i,i} \theb_i||_2^2$. 
We use $\theb_{j} (n)$ to emphasize the dependency on the phase vectors of other users in previous iteration.
The rest SCA update steps are the same.
Once all the phase updates are done, we calculate the effective end-to-end spatial signature update and conclude the second stage optimization.
The first stage remains the same.

\section{Numerical Results} 
\label{sec: exp}

\begin{figure}[t] %[htbp]
    \centering
    \includegraphics[scale=0.45]{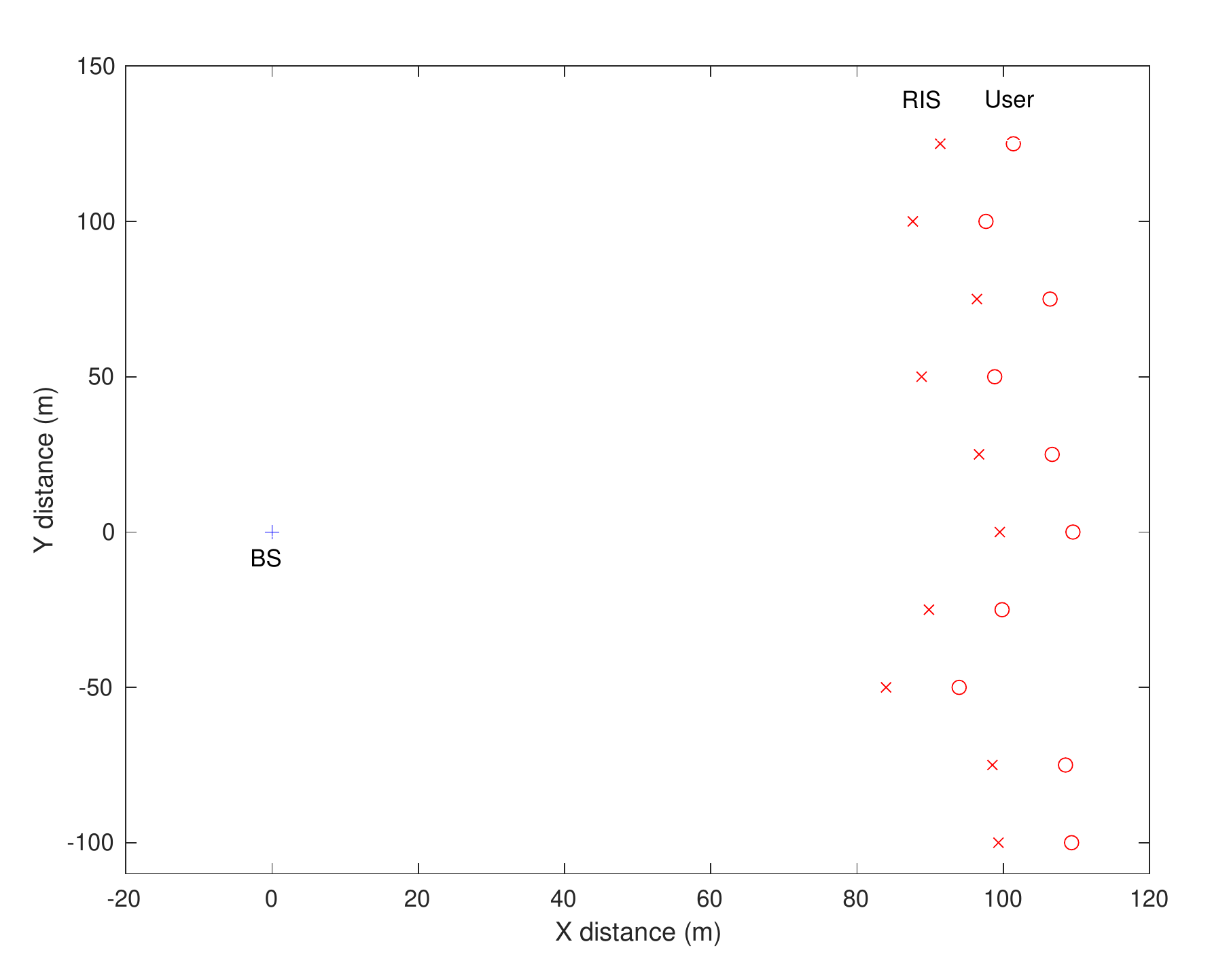}
    \caption{The simulated RISs-assisted communication system.}
    \label{fig:syssimu}
\end{figure}
We simulate a distributed RIS-assisted multiuser SIMO uplink communication system. 
Base station is equipped with $M=8$ antennas.  There are $N=10$ users and RISs, each RIS contains $K=100$ elements. The base station is located at (0m,0m), all the users are uniformly distributed in the x dimension in the range [-90m,110m], in the y dimension the distance between two adjacent users are set to 25m, as shown in Fig.~\ref{fig:syssimu}. Each RIS is located 10 meters ahead of its corresponding user. In the experiments, we set a common SINR target value $\gamma_i^* = 3.5 \approx 5 dB$ for all the users. The variance of AWGN noise $\sigma^2$ is $10^{-13}$. 
We use a path loss exponent $2$ for both user-RIS link and RIS-BS link.%, which are explained in section~\ref{sec: prelim}. 
The path loss exponent of BS-user direct link is set to $3$, so that we can provide comparisons with a system without the RIS assistance, though as per the model, our RIS-assisted system is designed not to rely on the direct links, i.e., they are negligible compared to the two hop RIS assisted link \cite{sanguinetti2012tutorial}.

We consider several baselines:
\begin{list}{\labelitemi}{\leftmargin=1em \itemindent=-0.5em \itemsep=.2em}
\item Baseline 1: No RISs in the system, only direct links.
\item Baseline 2: RIS-assisted system with random phase design.
\item Baseline 3 (Lower bound): RIS-assisted system without phase constraint, and least square solution can be used.
\end{list}

In simulations, we find that though \alg converges constantly, the convergence can be relatively slow, due to the several nested iterative algorithmic structure. 
In order to accelerate the convergence speed, instead of computing the optimal end-to-end spatial signature each time, we simply find one feasible solution by setting Lagrange multiplier to be zero and then normalize the resulting spatial signature to the maximum norm. 
Next, we use SCA to find optimal phase design as described in section~\ref{sec: prob}. These steps typically can speed up the convergence of algorithm significantly in practice.

\begin{figure} [t] %[htbp]
    \centering
    \includegraphics[scale=0.45]{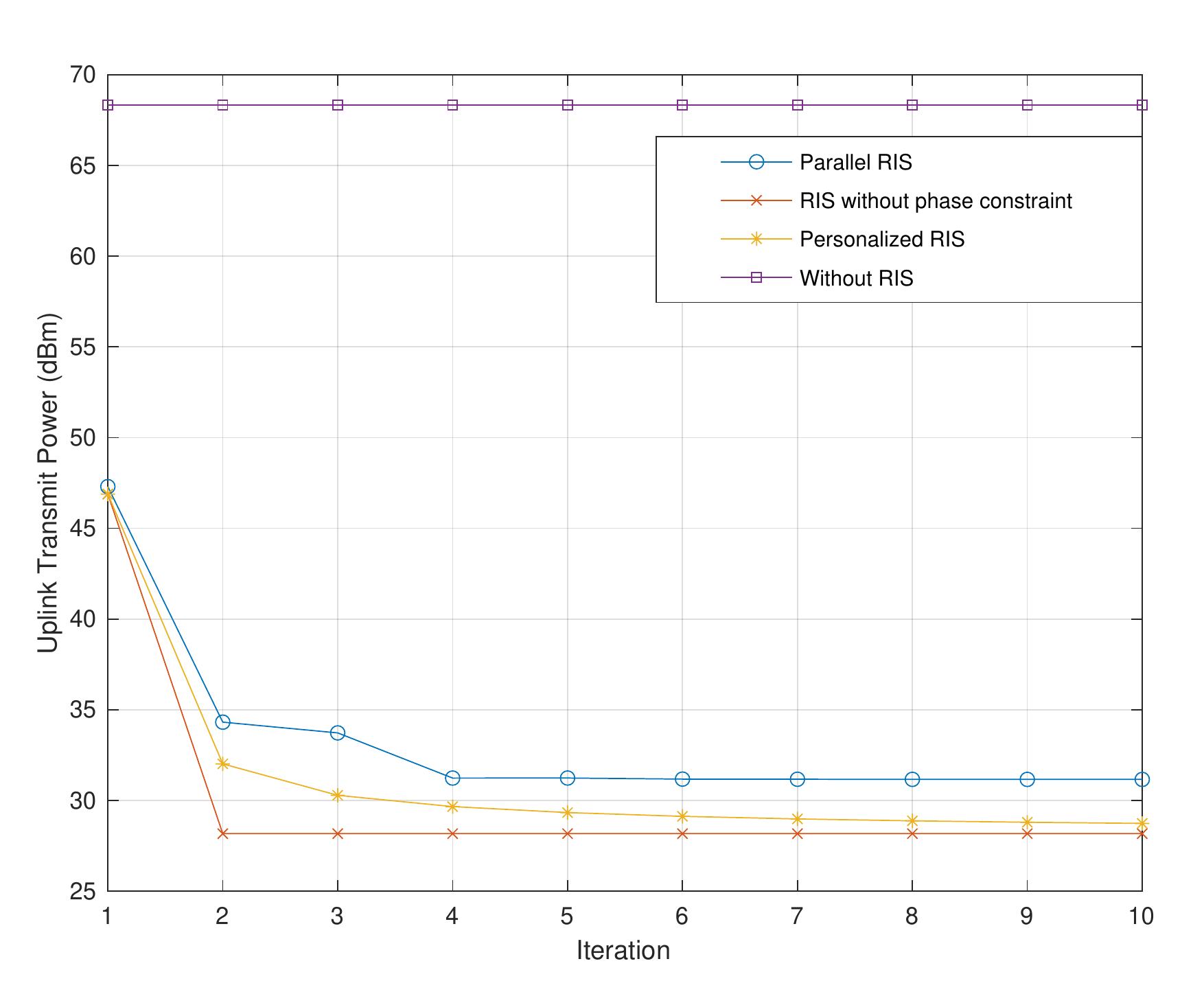}
    \caption{Total uplink transmit power.}
    \label{fig:comp4sys}
\end{figure}

In Fig.~\ref{fig:comp4sys}, the total uplink transmit power with respect to the outer loop iteration $t$ is shown. 
Note that even RIS assisted system with random phase (the first point of RIS systems) performs better than the one without RIS. This illustrates the advantage of applying RISs. 
We take two models into account, the first system without interference (which is represented as personalized RIS) and the second one with interference (which is represented as parallel RIS) as explained in section~\ref{sec: prelim}. 
For reference, the lower bound in total power is also plotted, which represents the RIS system without phase constraints. We observe that the cost of the physical constraints of the RIS, i.e., the phase constraint is not large, the proposed methods converge to the fixed power point that is close to the lower bound. Finally, including interference in system model leads to higher total power.%, as we need more transmit power to mitigate such interference.

\begin{figure} [t] %[htbp]
    \centering
    \includegraphics[scale=0.5]{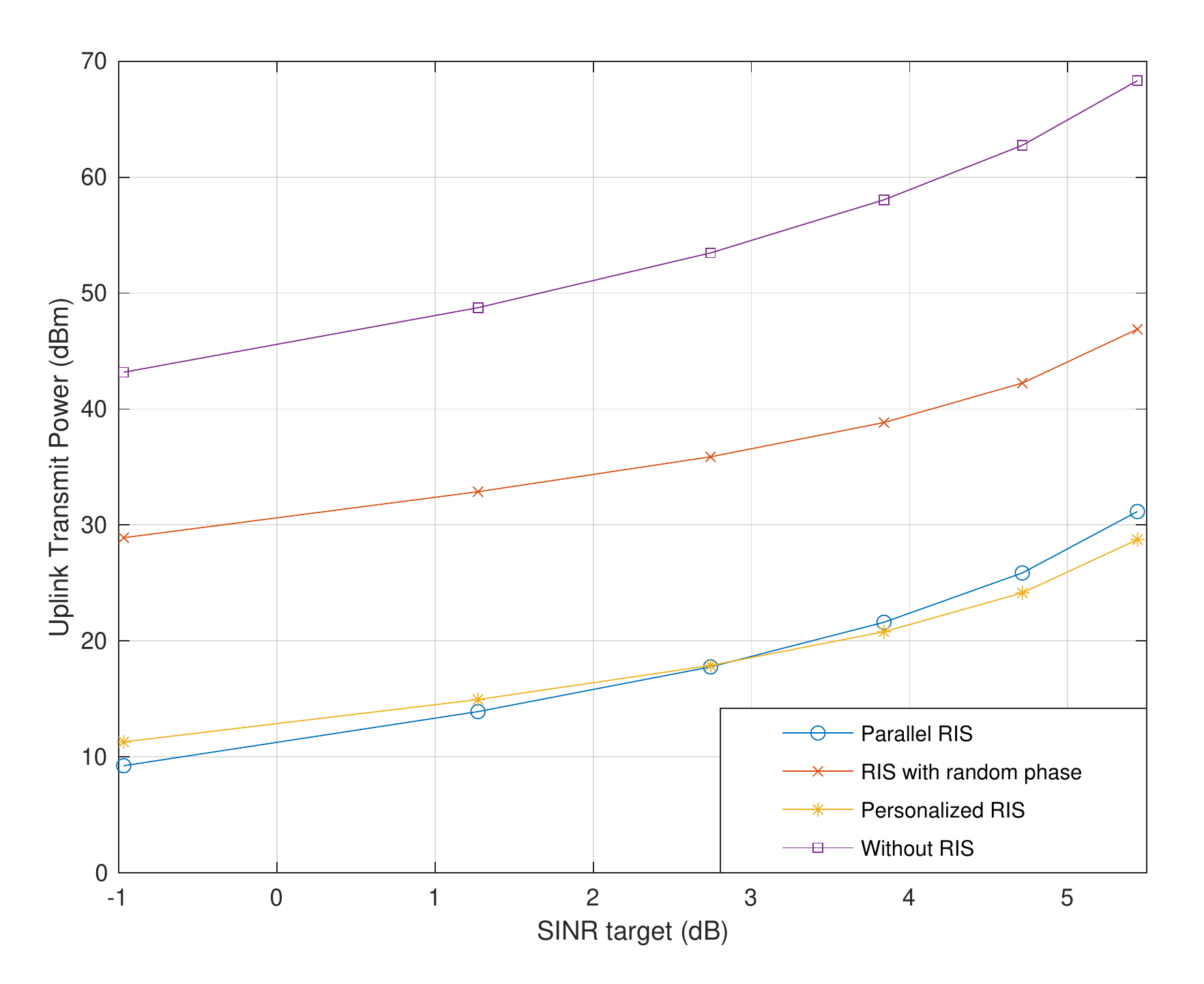}
    \caption{Total transmit power with different QoS requirements comparison.}
    \label{fig:comp4sinr}
\end{figure}

In Fig.~\ref{fig:comp4sinr}, we compare four systems under varying SINR targets.
With the increasing SINR requirements, the total transmit power increases as well for all set ups, and the without RISs expends the most power.
The proposed method results in less total power compared to the random phase design, which demonstrates its effectiveness.
One interesting phenomenon is that when SINR target is small, the parallel RIS model capitalizes on the collective paths from all users from all RISs and achieves slightly less power than the personal RIS-assisted set up. 
%This is because when quality of service requirement is low, the interference between user and other RISs actually acts as a positive role to help transmit signals.

\section{Conclusion} 
\label{sec: conclusion}

In this paper, we have developed a new iterative algorithm called \alg to jointly design transmit powers, receiver filters and RIS phases in order to minimize the total uplink transmit power under QoS constraints of an RIS-enhanced wireless system with as many RIS units as uplink users.
We have considered two distributed RIS assisted system models, namely personal RIS model and parallel RIS model, and provided the power, receiver filter and phase design updated for each user. We have applied successive convex approximation technique to deal with the non-convex constraint of phase shifts. Numerical results have verified that the proposed algorithm~\alg is effective in saving total transmit power when the direct links are weak. Future directions include exploring the impact of active RISs, discrete phase shifts, and phase-dependent reflection amplitude.

%\newpage
%\vspace{0.5in}
\bibliographystyle{IEEEtran}{}
\bibliography{BIB/CDMA, BIB/Optimization,BIB/Relay, BIB/RIS, BIB/RISpowercontrol}

\begin{thebibliography}{10}
\providecommand{\url}[1]{#1}
\csname url@samestyle\endcsname
\providecommand{\newblock}{\relax}
\providecommand{\bibinfo}[2]{#2}
\providecommand{\BIBentrySTDinterwordspacing}{\spaceskip=0pt\relax}
\providecommand{\BIBentryALTinterwordstretchfactor}{4}
\providecommand{\BIBentryALTinterwordspacing}{\spaceskip=\fontdimen2\font plus
\BIBentryALTinterwordstretchfactor\fontdimen3\font minus
  \fontdimen4\font\relax}
\providecommand{\BIBforeignlanguage}[2]{{%
\expandafter\ifx\csname l@#1\endcsname\relax
\typeout{** WARNING: IEEEtran.bst: No hyphenation pattern has been}%
\typeout{** loaded for the language `#1'. Using the pattern for}%
\typeout{** the default language instead.}%
\else
\language=\csname l@#1\endcsname
\fi
#2}}
\providecommand{\BIBdecl}{\relax}
\BIBdecl

\bibitem{saad2019vision}
W.~Saad, M.~Bennis, and M.~Chen, ``{A Vision of 6G Wireless Systems:
  Applications, Trends, Technologies, and Open Research Problems},'' \emph{IEEE
  network}, vol.~34, no.~3, pp. 134--142, 2019.

\bibitem{wu2019intelligent}
Q.~Wu and R.~Zhang, ``{Intelligent Reflecting Surface Enhanced Wireless Network
  via Joint Active and Passive Beamforming},'' \emph{IEEE Transactions on
  Wireless Communications}, vol.~18, no.~11, pp. 5394--5409, 2019.

\bibitem{wu2019towards}
------, ``{Towards Smart and Reconfigurable Environment: Intelligent Reflecting
  Surface Aided Wireless Network},'' \emph{IEEE Communications Magazine},
  vol.~58, no.~1, pp. 106--112, 2019.

\bibitem{wu2021intelligent}
Q.~Wu \emph{et~al.}, ``{Intelligent Reflecting Surface-Aided Wireless
  Communications: A Tutorial},'' \emph{IEEE Transactions on Communications},
  vol.~69, no.~5, pp. 3313--3351, 2021.

\bibitem{hu2018beyond}
S.~Hu, F.~Rusek, and O.~Edfors, ``{Beyond Massive MIMO: The Potential of Data
  Transmission with Large Intelligent Surfaces},'' \emph{IEEE Transactions on
  Signal Processing}, vol.~66, no.~10, pp. 2746--2758, 2018.

\bibitem{chen2008distributed}
M.~Chen, S.~Serbetli, and A.~Yener, ``{Distributed Power Allocation Strategies
  for Parallel Relay Networks},'' \emph{IEEE Transactions on Wireless
  Communications}, vol.~7, no.~2, pp. 552--561, 2008.

\bibitem{zhao2020energy}
J.~Zhao \emph{et~al.}, ``{Energy Efficient Full-Duplex Communication Systems
  with Reconfigurable Intelligent Surface},'' in \emph{2020 IEEE 92nd Vehicular
  Technology Conference (VTC2020-Fall)}, 2020, pp. 1--5.

\bibitem{wu2020jointpower}
Q.~Wu and R.~Zhang, ``{Joint Active and Passive Beamforming Optimization for
  Intelligent Reflecting Surface Assisted SWIPT under QoS Constraints},''
  \emph{IEEE Journal on Selected Areas in Communications}, vol.~38, no.~8, pp.
  1735--1748, 2020.

\bibitem{guo2021joint}
B.~Guo, R.~Li, and M.~Tao, ``{Joint Design of Hybrid Beamforming and Phase
  Shifts in RIS-Aided mmWave Communication Systems},'' in \emph{2021 IEEE
  Wireless Communications and Networking Conference (WCNC)}.\hskip 1em plus
  0.5em minus 0.4em\relax IEEE, 2021, pp. 1--6.

\bibitem{fu2021reconfigurable}
M.~Fu \emph{et~al.}, ``{Reconfigurable Intelligent Surface Empowered Downlink
  Non-orthogonal Multiple Access},'' \emph{IEEE Transactions on
  Communications}, vol.~69, no.~6, pp. 3802--3817, 2021.

\bibitem{zhou2021joint}
Y.~Zhou \emph{et~al.}, ``{Joint Sensor Selection, Beamforming and Phase Control
  in Reconfigurable Intelligent Surface Aided IoT Fusion Networks},''
  \emph{IEEE Wireless Communications Letters}, 2021.

\bibitem{wang2020power}
H.~Wang \emph{et~al.}, ``{Power Minimization for Two-Cell IRS-Aided NOMA
  Systems with Joint Detection},'' \emph{IEEE Communications Letters}, vol.~25,
  no.~5, pp. 1635--1639, 2020.

\bibitem{wang2022simultaneous}
Y.~Wang \emph{et~al.}, ``{Simultaneous Transmission and Reflection
  Reconfigurable Intelligent Surface Assisted Full-Duplex Communications},''
  \emph{arXiv preprint arXiv:2203.05411}, 2022.

\bibitem{cao2021delay}
Y.~Cao \emph{et~al.}, ``{Delay-constrained Joint Power Control, User Detection
  and Passive Beamforming in Intelligent Reflecting Surface-assisted Uplink
  mmWave System},'' \emph{IEEE Transactions on Cognitive Communications and
  Networking}, vol.~7, no.~2, pp. 482--495, 2021.

\bibitem{wu2022energy}
J.~Wu, S.~Kim, and B.~Shim, ``{Energy-Efficient Power Control and Beamforming
  for Reconfigurable Intelligent Surface-Aided Uplink IoT Networks},''
  \emph{arXiv preprint arXiv:2203.09090}, 2022.

\bibitem{ma2021power}
H.~Ma and H.~Wang, ``{Power Minimization Transmission Design for IRS-Assisted
  Uplink NOMA Systems},'' in \emph{2021 IEEE 94th Vehicular Technology
  Conference (VTC2021-Fall)}.\hskip 1em plus 0.5em minus 0.4em\relax IEEE,
  2021, pp. 1--4.

\bibitem{liu2020intelligent}
Y.~Liu \emph{et~al.}, ``{Intelligent Reflecting Surface Aided MISO Uplink
  Communication Network: Feasibility and Power Minimization for Perfect and
  Imperfect CSI},'' \emph{IEEE Transactions on Communications}, vol.~69, no.~3,
  pp. 1975--1989, 2020.

\bibitem{pan2021overview}
C.~Pan \emph{et~al.}, ``{An Overview of Signal Processing Techniques for
  RIS/IRS-aided Wireless Systems},'' \emph{arXiv preprint arXiv:2112.05989},
  2021.

\bibitem{yener2001inter}
A.~Yener, R.~D. Yates, and S.~Ulukus, ``{Interference Management for CDMA
  Systems through Power Control, Multiuser Detection, and Beamforming},''
  \emph{IEEE Transactions on Communications}, vol.~49, no.~7, pp. 1227--1239,
  2001.

\bibitem{erpek2021autoencoder}
T.~Erpek \emph{et~al.}, ``{Autoencoder-based Communications with Reconfigurable
  Intelligent Surfaces},'' \emph{IEEE International Symposium on Dynamic
  Spectrum Access Networks (DySPAN)}, pp. 242--247, 2021.

\bibitem{xu2013block}
Y.~Xu and W.~Yin, ``{A Block Coordinate Descent Method for Regularized
  Multiconvex Optimization with Applications to Nonnegative Tensor
  Factorization and Completion},'' \emph{SIAM Journal on imaging sciences},
  vol.~6, no.~3, pp. 1758--1789, 2013.

\bibitem{ulukus1998adaptive}
S.~Ulukus and R.~D. Yates, ``{Adaptive Power Control and MMSE Interference
  Suppression},'' \emph{Wireless Networks}, vol.~4, no.~6, pp. 489--496, 1998.

\bibitem{ulukus2004iterative}
S.~Ulukus and A.~Yener, ``{Iterative Transmitter and Receiver Optimization for
  CDMA Networks},'' \emph{IEEE Transactions on Wireless Communications},
  vol.~3, no.~6, pp. 1879--1884, 2004.

\bibitem{scutari2013decomposition}
G.~Scutari \emph{et~al.}, ``{Decomposition by Partial Linearization: Parallel
  Optimization of Multi-Agent Systems},'' \emph{IEEE Transactions on Signal
  Processing}, vol.~62, no.~3, pp. 641--656, 2013.

\bibitem{guo2020weighted}
H.~Guo \emph{et~al.}, ``{Weighted Sum-Rate Maximization for Reconfigurable
  Intelligent Surface Aided Wireless Networks},'' \emph{IEEE Transactions on
  Wireless Communications}, vol.~19, no.~5, pp. 3064--3076, 2020.

\bibitem{sanguinetti2012tutorial}
L.~Sanguinetti, A.~A. D'Amico, and Y.~Rong, ``{A Tutorial on the Optimization
  of Amplify-and-Forward MIMO Relay Systems},'' \emph{IEEE Journal on Selected
  Areas in Communications}, vol.~30, no.~8, pp. 1331--1346, 2012.

\end{thebibliography}

%\onecolumn
%\input{proof.tex}

\end{document}